\documentclass[]{aa}
\usepackage{graphicx}
\begin{document}
\title{An XMM-Newton view of the serendipitous sources in the PKS0312-770 field}
\author{D H Lumb \inst{1}
	\and M Guainazzi \inst{2}
	\and P Gondoin \inst{1} }
\institute{XMM-Newton SOC, Space Science Dept., ESA, ESTEC, 2200AG
Noordwijk, Netherlands \and XMM-Newton SOC, Space Science Dept., ESA, Villafranca Satellite Tracking Station, 28080 Madrid }
\thanks{This work is based on observations made with the XMM-Newton, an
ESA science mission with instruments and contributions directly funded by
ESA member states and the USA (NASA).}
\offprints{D Lumb@astro.estec.esa.nl}

\date{Received date / Accepted date}
\titlerunning{XMM follow-up of CHANDRA}
\maketitle
\abstract{We describe an XMM-Newton observation of the PKS0312-770 field, 
which facilitates the spectral analysis of serendipitous sources  previously 
detected by CHANDRA. The combination of larger effective area and longer exposure 
duration allows a significant increase in detected photons, and a lower limit in source 
detection sensitivity. In particular the
hard X-ray normal galaxy unveiled by Fiore et al (2000) is most likely
explained as a moderately absorbed (N$_{H} \sim$ 10$^{22}$cm$^{-2}$) AGN. 
We detect 52 sources (45 previously unreported) at a limiting flux of
$\sim$2 10$^{-15}$ergs cm$^{-2}$ s$^{-1}$ in the 0.5-2keV band. The LogN-LogS 
curve is consistent with that derived from  by XMM-Newton observations of the Lockman Hole field.
 The flux determinations allow to check for any
inconsistency between the calibrations of  the two observatories, which
is discussed.} 

\section{Introduction}
 Much of the soft X-ray background was resolved by ROSAT (in the Lockman Hole for example, Hasinger et al \cite{Hasinger98}),
where the majority of sources have been identified with AGNs, although a significant fraction of obscured, 
 hard sources - probably type II AGNs (Lehmann et al \cite{Lehmann}) - were also identified.
The characterisation of these faintest X-ray source populations is being revolutionised 
by the observations made by the CHANDRA and XMM-Newton Observatories.  Their 
capabilities are somewhat complementary: the unprecedented CHANDRA 
angular resolution (Van Speybroek et al \cite{vanSpeybroek}) allows for 
negligible background and ultimate source detection sensitivity; while 
the XMM-Newton telescopes (Jansen et al \cite{Jansen}) offer the largest ever 
focussed effective area for unmatched photon gathering power.

The first deep field observations performed by these observatories (Giacconi et al
\cite{Giacconi}; Hasinger et al \cite{Hasinger}; Hornschemeier et al \cite{Hornschemeier}  and Mushotzky
et al \cite{Mushotzky}) have confirmed these promises. CHANDRA
observations to  a source limiting sensitivity of $\sim$
2$\times$10$^{-16}$ ergs cm$^{-2}$ s$^{-1}$ (0.5-2keV) resolved  about 80\% of the 
background,  and found many hard spectra at faint levels which helps 
to resolve the ``spectral paradox'' of the difference between the spectrum 
of the background and the spectrum of bright AGN. 
Hornschemeier et al \cite{Hornschemeier} also note 
 an increase in proportion of normal galaxies at flux levels
$\leq$3$\times$10$^{-16}$ ergs cm$^{-2}$ s$^{-1}$.
XMM-Newton pushed the limits further than CHANDRA in the 5-10keV band,
reaching 2.4 $\times$10$^{-15}$ ergs
cm$^{-2}$ s$^{-1}$ (Hasinger et al \cite{Hasinger}).  The optical follow-up of all these deep
fields is still subject to extensive effort. 

Relative flux normalisations of the XMM-Newton 
and CHANDRA observatories are as yet little explored. This normalisation  has
a significant impact on the studies of the source populations comprising the X-ray
background, as well as on the analysis of the Sunayev-Zeldovich effect and for
measurements of source variability over long temporal baselines. In the present
study we compared CHANDRA and XMM-Newton data from this PKS0312-770 field
to examine this normalisation calibration.

The deep fields mentioned above, have been thoroughly studied at 
all wavelengths, so that the identification of many objects is secure. Nevertheless, 
the latest harvest of fainter objects is overwhelmingly in the very red and 
faint end of the optical population of galaxies, rendering them difficult or 
impossible to analyse spectroscopically with even 8 - 10m class telescopes. 
To learn more about their nature and evolution requires the photometric estimate 
of their redshifts and/or more spectral classification of their X-ray properties. In this observation, which  is more typical of Guest Observer target observations, 
we show the distribution of hardness ratios that could help to identify
peculiar sources for follow-up.

\section{The Observation}
During the calibration phase of CHANDRA, the field of AGN \object{PKS0312-770}
was observed as part of the mirror point spread function determinations.
Fiore et al (\cite{Fiore}) described the detection of 6 sources from a 2-10keV 
image made by the 16 arcmin square ACIS-I detector. They then observed these 
objects  with the ESO 3.6m EFOSC2 spectrograph.  The identifications suggested 
4 broad-line quasars, one quasar with possible moderate obscuration, and
an apparently normal galaxy. Their X-ray analysis was limited by the low
count rates, to   $^{(0.5-2)}$/$_{(2-10)}$ keV  hardness ratios only, and flux
estimates based on an assumed photon power law spectral model. Except for the
normal galaxy, the hardness ratios were consistent with a power law  absorbed 
by the Galactic column. For the normal galaxy, the CHANDRA data did  not  allow  
Fiore  et  al to discriminate between (for example)  an obscured AGN  hypothesis or  scenarios 
involving beamed continuum emission and a flat power  law  Advection Dominated Accretion Flow
models.

PKS0312-770 was observed during the XMM-Newton calibration period to 
characterise the XMM mirror PSF. The observation 
was performed on 2000-03-31 at UT 14.00h. The three EPIC cameras 
(Str\"uder et al \cite{Struder}, Turner et al \cite{Turner}) were operated in a 
full frame readout mode, offering a $\sim$30 arcminute field of view. An optical blocking 
filter  with so-called THICK  aluminium layer (2000 \AA ~of Aluminium)  was employed. 

The data were reduced using the XMM-Newton Science Analysis Sub-system
(SASv5.0). Trials were made for a selection to discount periods of high
background, arising from intense soft proton fluxes generated in
magnetospheric reconnection events, and focussed by the mirrors. In a
number of other observations, we found that acceptable rejection was attained
by defining {\em Good Time Intervals} where the integrated flux above 10keV in energy 
was $\leq$ 2 (1) counts s$^{-1}$ in EPIC PN (MOS). Above this energy the flux is 
dominated by particles, rather than by X-rays. The resulting clean exposure
time after this selection was  about 27ks, or $\sim$ 80\% longer than the 
corresponding CHANDRA exposure, and was used for  maximum sensitivity  in the
detection of faint point sources.  Using a range of background 
rejection thresholds, we determined that in this particular observation the 
background proton flares were relatively modest, 
and we pursued the spectral analysis for all objects of the Fiore et al \cite{Fiore}  study with data
extracted from the {\em full} observation duration, in order to maximise
the photon counts.  

\begin{figure}
\begin{center}
\resizebox{\hsize}{!}{\includegraphics{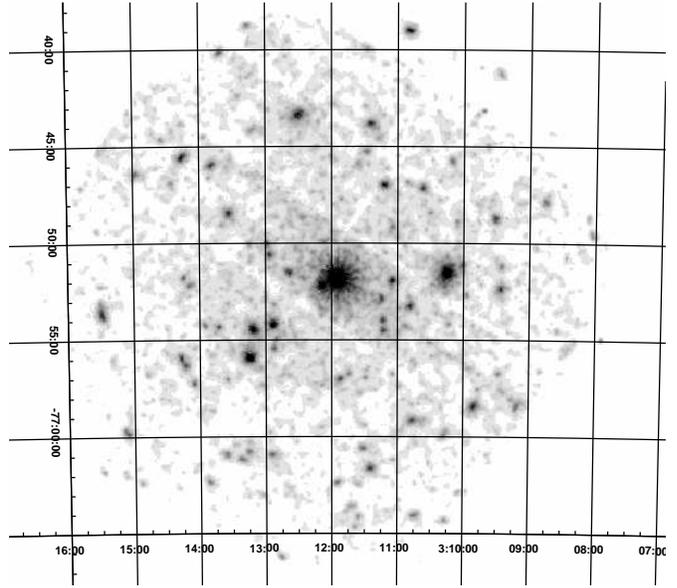}}
\caption{0.5-2keV merged image of the PKS0312-770 field. All 3 EPIC cameras, 
logarithmic scaling. A 4 arcsecs Gaussian smoothing has been applied }
\label{fig:softimage}
\end{center}
\end{figure}

\section{Point Source Detections}
Fig.~\ref{fig:softimage}  shows the merged image of all 3  EPIC
cameras, in the 0.5-2keV band, following a Gaussian smoothing of 4arcseconds.

 The
bright central target source was within 2 arcseconds of the requested pointing,
 and its centroid did not move significantly throughout the
observation. The early calibration and performance/verification phase, 
demonstrated that the achieved pointing {\em measurement } accuracy is 
typically 4 arcseconds r.m.s.. Our astrometric discrepancy is therefore well
within expectations

\begin{figure}
\resizebox{\hsize}{!}{\includegraphics{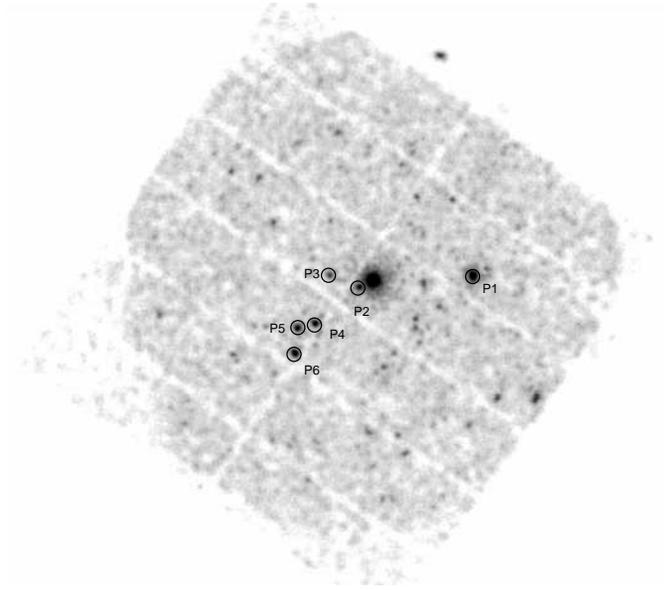}}
\caption{2 - 10 keV merged image of the PKS0312-770 field. All 3 EPIC cameras, 
logarithmic scaling. Circled and annotated are the serendipitous sources P1--P6 of Fiore
 et al \cite{Fiore}}
\label{fig:hardimage}
\end{figure}

 Marked on the image of Fig.~\ref{fig:hardimage} (in 2 - 10 keV) are the
 location  of  the  sources  P1  - P6 noted by Fiore et al	\cite{Fiore}. They were 
each found within a 3 arcsec radius of the reported CHANDRA locations. The average
displacement from CHANDRA-reported locations is -1 arcsecond in RA and
-1.5 arcseconds in Dec, with a possible field  rotation ($\ll$0.1 degree)accounting for
part of this discrepancy. Such a rotation could degrade position
locations for objects at the edge of our field to $\sim$5 - 6
arcseconds.

Table 1 tabulates the sources which are detected in the 0.5-2
and 2-10keV bands. This source detection list was established from the
reduced low background
(27ks) portion of the exposure. The XMM-SAS task $EBOXDETECT$ was used to
perform a   sliding box cell detection with local background subtraction, of both bands
simultaneously.
A detection  threshold  of  5$\sigma$ was used.  We detect 52 sources in the soft
band, of which 47 are detected in the hard band. 45 out of the 52 objects are
previously unreported.  For  the  newly detected
sources, we estimate  their   0.5-2keV fluxes assuming a simple
 power law spectrum ($\Gamma$=1.7, N$_{H}$=8 10 $^{20}$). The fluxes of the 
first seven objects, already
known from CHANDRA, were derived by spectral fitting, as described in
Section~\ref{sec:spectra}.

\begin{figure}
\resizebox{\hsize}{!}{\includegraphics{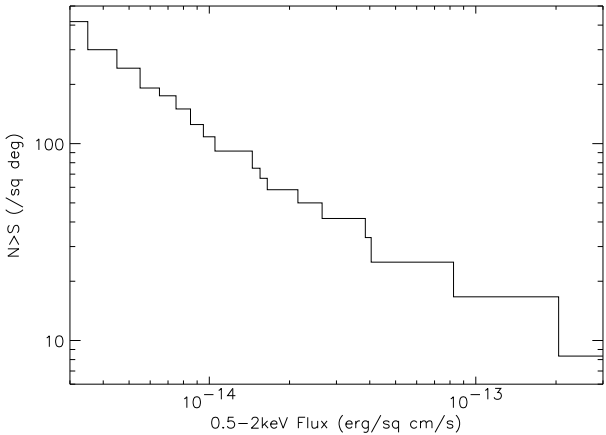}}
\caption{0.5 - 2 keV cummulative LogN-LogS curve for the PKS0312-770
field. } \label{fig:lognlogs}
\end{figure}

\begin{table*}
\caption{List of sources with significance $\geq$5$\sigma$, found using a sliding cell detection algorithm} 
\begin{tabular}{llllll}

Source ID& RA J2000 & Dec J2000&(0.5-2kev)& (2-10keV)&Hardness\\
&&&\multicolumn{2}{c}{Flux 10$^{-15}$ ergs cm$^{-2}$ s$^{-1}$}&Ratio\\
PKS0312-770 & 3 11 54.9 & -76 51 49.6 &   1300&2496&    -0.44\\ 
P1 & 3 10 15.3 & -76 51 32.4 &   205 &350 &    -0.35\\
P2 & 3 12  8.7 & -76 52 11.9 &   39  & 33 &    -0.58\\
P3 & 3 12 38.7 & -76 51 31.4 &   6   &  27&    0.01\\
P4 & 3 12 53.6 & -76 54 13.3 &   27  &  68&    -0.20\\
P5 & 3 13 11.7 & -76 54 28.9 &   41  &  65&    -0.34\\
P6 & 3 13 14.3 & -76 55 54.4 &   83  &  12&    -0.40\\ 
XMMU J031049.6-763901 & 3 10 49.6 & -76 39  1.5 &   22  &  55&    -0.16\\
XMMU J031245.7-770616 & 3 12 45.7 & -77  6 16.0 &   17  & 29 &    -0.36\\
XMMU J031230.6-764323 & 3 12 30.6 & -76 43 23.5 &   16  & 22 &    -0.60\\
XMMU J031529.1-765340 & 3 15 29.1 & -76 53 40.8 &   15  & 26 &    -0.41\\
XMMU J031416.1-764535 & 3 14 16.1 & -76 45 35.1 &   15  & 23 &    -0.49\\
XMMU J031124.4-764349 & 3 11 24.4 & -76 43 49.4 &   11  & 1.2&    -0.73\\
XMMU J030951.1-765825 & 3  9 51.1 & -76 58 25.3 &   11  & 41 &    -0.13\\
XMMU J031105.1-765156 & 3 11  5.1 & -76 51 56.7 &   10  & 22 &    -0.37\\
XMMU J031340.8-764009 & 3 13 40.8 & -76 40  9.7 &   9.3 & 13 &    -0.30\\
XMMU J031334.1-764827 & 3 13 34.1 & -76 48 27.5 &   8.8 & 19 &    -0.32\\
XMMU J030931.6-764845 & 3  9 31.6 & -76 48 45.7 &   8.7 & 7.6&    -0.50\\
XMMU J031348.8-764557 & 3 13 48.8 & -76 45 57.5 &   8.6 & 27 &    -0.17\\
XMMU J030927.0-765224 & 3  9 27.0 & -76 52 24.2 &   8.0 & 2.4&    -0.79\\
XMMU J031124.4-770137 & 3 11 24.4 & -77  1 37.9 &   7.2 & 15 &    -0.26\\
XMMU J031049.1-765316 & 3 10 49.1 & -76 53 16.2 &   7.1 & 16 &    -0.33\\
XMMU J031037.1-764710 & 3 10 37.1 & -76 47 10.7 &   6.7 & 20 &    -0.08\\
XMMU J031112.5-764659 & 3 11 12.5 & -76 46 59.7 &   6.3 & 7.4&    -0.47\\
XMMU J031416.3-765558 & 3 14 16.3 & -76 55 58.0 &   5.9 & 13 &    -0.29\\
XMMU J031011.0-764546 & 3 10 11.0 & -76 45 46.3 &   5.2 & 2.2&    -0.72\\
XMMU J031320.2-770110 & 3 13 20.2 & -77  1 10.9 &   5.2 & 8.1&    -0.21\\
XMMU J031256.6-765036 & 3 12 56.6 & -76 50 36.2 &   5.1 & 2.7&    -0.68\\
XMMU J030911.9-765824 & 3  9 11.9 & -76 58 24.9 &   5.1 & 8.1&    0.43\\
XMMU J031114.5-765252 & 3 11 14.5 & -76 52 52.8 &   4.9 & 1.6&    -0.76\\
XMMU J031148.9-770222 & 3 11 48.9 & -77  2 22.5 &   4.9 & 4.1&    -0.47\\
XMMU J030803.3-764938 & 3  8  3.3 & -76 49 38.3 &   4.7 & 10 &    -0.21\\
XMMU J031113.6-765358 & 3 11 13.6 & -76 53 58.4 &   4.7 & 15 &    -0.11\\
XMMU J031113.3-765430 & 3 11 13.3 & -76 54 30.7 &   4.4 & 12 &    -0.22\\
XMMU J030928.6-765642 & 3  9 28.6 & -76 56 42.1 &   4.4 & - &    -1.00\\
XMMU J030925.7-765109 & 3  9 25.7 & -76 51  9.0 &   4.1 & 8.5&    -0.23\\
XMMU J031315.0-770056 & 3 13 15.0 & -77  0 56.0 &   3.9 & 4.6&    -0.59\\
XMMU J031045.3-770405 & 3 10 45.3 & -77  4  5.1 &   3.9 & 7.7&    -0.29\\
XMMU J031047.3-765909 & 3 10 47.3 & -76 59  9.6 &   3.8 &  - &    -1.00\\
XMMU J031415.3-765716 & 3 14 15.3 & -76 57 16.6 &   3.8 & - &    -1.00\\
XMMU J031152.3-765701 & 3 11 52.3 & -76 57  1.7 &   3.7 & 4.4&    -0.56\\
XMMU J031342.5-765421 & 3 13 42.5 & -76 54 21.5 &   3.6 & 8.9&    -0.14\\
XMMU J031128.0-764516 & 3 11 28.0 & -76 45 16.4 &   3.6 & 1.6&    -0.47\\
XMMU J031412.6-765619 & 3 14 12.6 & -76 56 19.6 &   3.5 & 9.1&    -0.24\\
XMMU J031154.7-770221 & 3 11 54.7 & -77  2 21.7 &   3.5 & 11 &    -0.01\\
XMMU J031315.6-770047 & 3 13 15.6 & -77  0 47.1 &   3.5 & 5.8&    -0.27\\
XMMU J031259.0-765001 & 3 12 59.0 & -76 50  1.3 &   3.4 &  - &    -1.00\\
XMMU J031412.5-765154 & 3 14 12.5 & -76 51 54.0 &   3.4 & 3.3&    -0.38\\
XMMU J031010.1-765956 & 3 10 10.1 & -76 59 56.3 &   3.3 & - &    -1.00\\
XMMU J031001.5-765107 & 3 10 1.50 & -76 51  7.9 &   3.2 & 12 &    -0.09\\
XMMU J031131.4-770036 & 3 11 31.4 & -77  0 36.2 &   2.6 & 9.9&    0.04\\
XMMU J031252.6-765525 & 3 12 52.6 & -76 55 25.7 &   2.2 & 2.3&    -0.42\\
 \label{tab:sourcelist}

\end{tabular}

 \end{table*}
Fig.~\ref{fig:lognlogs} shows the cummulative LogN-LogS distribution
extracted from the 0.5-2keV detections, with the fluxes based on the same simple 
power-law model of emission spectrum. In the flux range 10$^{-15}$ --
10$^{-14}$  ergs cm$^{-2}$ s$^{-1}$ our source counts are consistent 
 with those presented in Hasinger et al (\cite{Hasinger}), given the
 different values of Galactic absorption in the two fields.

\begin{figure}
\resizebox{\hsize}{!}{\includegraphics{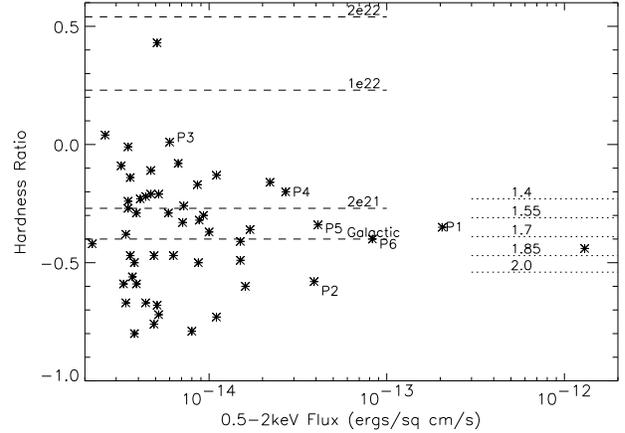}}
\caption{Hardness ratio versus 0.5 - 2 keV flux. Location of sources mentioned in the text are noted.  The dotted lines display the HR for different power laws when the absorption is fixed to the Galactic value.  The dashed	 lines are the equivalent loci for $\Gamma$=1.7 with different absorption.} \label{fig:hr}
\end{figure}

In Fig.~\ref{fig:hr} we plot the hardness ratios 
(C$_{2-10}$-C$_{.5-2}$)/(C$_{2-10}$+C$_{.5-2}$) as a
function of soft-band flux. This classification allows for follow up of 
source populations, for example selecting particularly hard / absorbed spectra
representing type-2 hidden AGN. Dashed horizontal
lines provide the location of different intrinsic absorption, when $\Gamma$=1.7. Likewise
the dotted lines reveal the location for hardness ratio at a given power law slope
for the Galactic absorption only. The hardest source in the plot is the target XMMU
J030911.9-765824, whose spectrum is consistent with absorption of $\geq$
10$^{22}$cm$^{-2}$. It should be noted that care has to be taken in interpreting
such plots, as particularly at the edges of the field of view, it is possible for
either the PN or MOS coverage to be totally lost (see for example the hard source to
N of PKS0312-770 at the field edge; it is covered only by the MOS fields of view).
The uniform energy conversion factors applied for band ratios can be affected by 
energy-dependent vignetting factors which vary {\em within} coarse energy bands.

\section{Spectral Analysis}\label{sec:spectra}
Standard data selections in SAS comprise the removal of bad pixels,
and pixels adjacent to  CCD borders, of bad CCD readout frames and of spurious noise events. 
For the  PN camera, only events contained  within one or two pixels have an accurate
spectral calibration and thus the remainder were excluded from the
spectral fitting. For the MOS camera the canonical selection is for events 
classified on-board with a ``PATTERN'' identifier. 

For each source, an extraction radius
of 45 arcsec        was used, unless limited by an artefact such as a
CCD chip boundary. The associated encircled energy fraction is $\geq$90\%.
It is weakly energy   dependent, but accounted for in the generation of the
appropriate response functions.   An additional  correction for the vignetting
was applied. For the PN camera this is a function of radius
within the field of view. 
At a field angle of 10 arcminutes vignetting becomes energy dependent
only  above 5keV. None of the 6 brightest sources reported from CHANDRA,
is further than 6 arcminutes off-axis and thus with a vignetting value of
less than 75\%. Thus
any remnant {\em mis-calibration} will be a negligible factor for
spectral fitting. 

In the case of the MOS cameras the presence of the Reflection Grating Arrays, 
located behind the mirror modules  generate an additional
obscuration factor. This obscuration is almost independent of
energy, but  strongly depends on the azimuthal angle in the EPIC MOS focal plane. 
This feature is corrected in the source-specific generated response
distributions with an accuracy better than $\sim$3\% at a 6 arcminute
off-axis angle. The residual effects are likely to affect only the
calculated flux, and not the fitted power laws or absorption values. 

A background region was extracted  as an annulus around each source, and within 
the same CCD. An exception was for source P2, which is within the wings of 
the bright on-axis target. For this source, a region of identical size to the P2
extraction circle was taken, but at the same distance from the PKS0312-770 location,
only moved in azimuth around the central target. This provides a comparable
amount of contamination in the background region and in the source region.

For each source we performed a combined spectral fit with the
data of the PN and both MOS cameras, using XSPEC v11.0.1.	To compare
with the data of Fiore et al ( power law with galactic absorption) we
used a similar description, but allowed an additional absorption component at the
redshift of the targets. In most
cases there was no strong evidence for absorption in excess of the
galactic column density (Table 2).

\begin{figure}
\resizebox{\hsize}{!}{\includegraphics{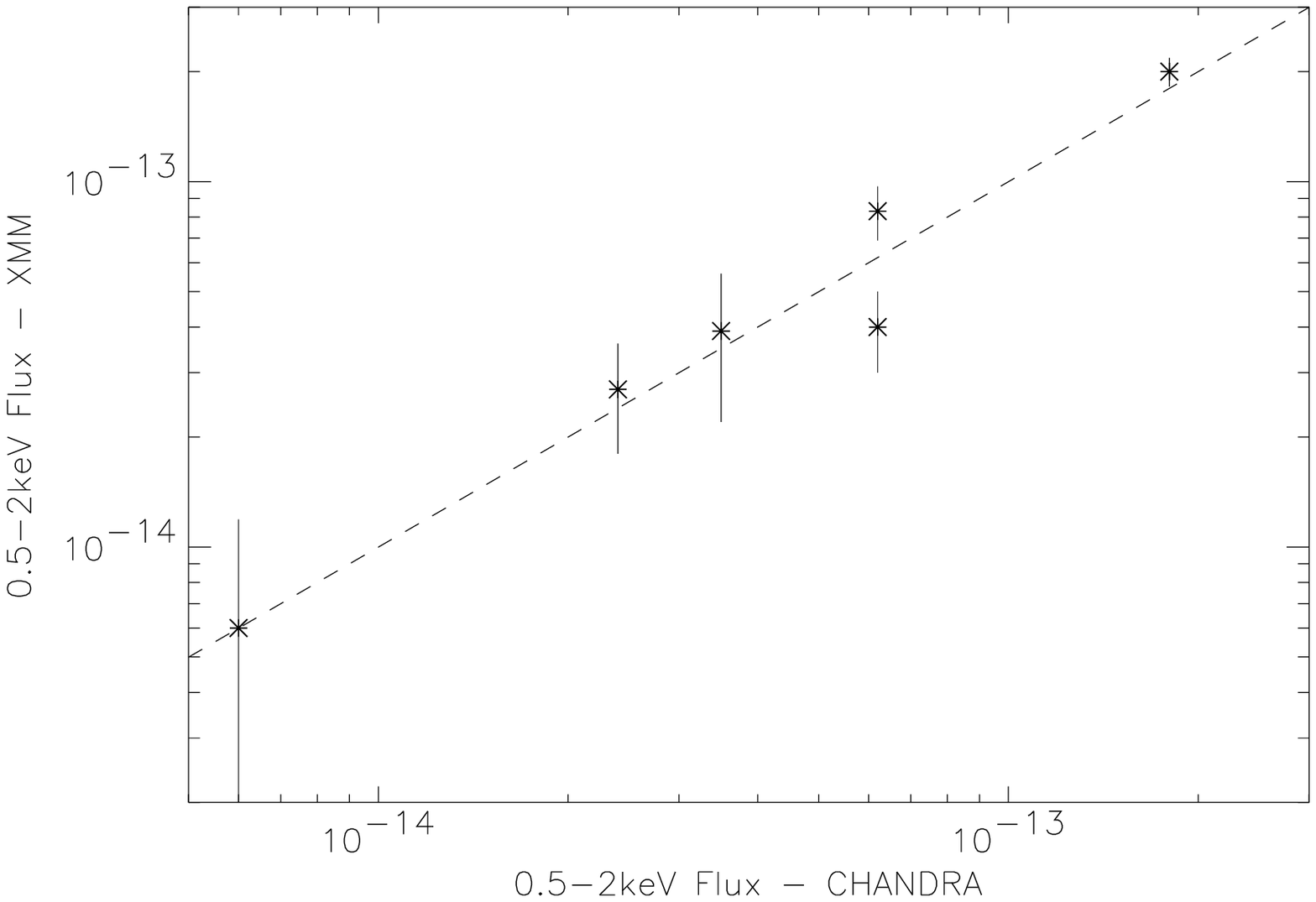}}
\caption{Comparison of estimated soft band fluxes in XMM-Newton and CHANDRA (from Fiore et al \cite{Fiore}). The XMM error bars are dominated by those of the spectral fitting parameters. No estimates were given for CHANDRA }
\label{fig:fluxcompa}
\end{figure}

\begin{figure}
\resizebox{\hsize}{!}{\includegraphics{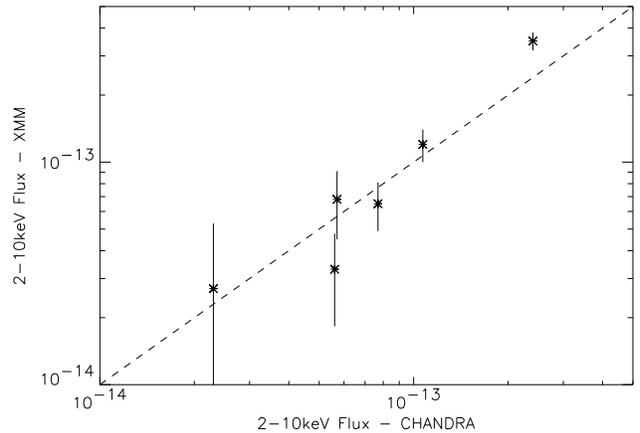}}
\caption{Comparison of estimated hard band fluxes in XMM-Newton and CHANDRA (from Fiore et al \cite{Fiore}). The XMM error bars are dominated by those of the spectral fitting parameters. No estimates were given for CHANDRA }
\label{fig:fluxcompb}
\end{figure}

Following these spectral
analyses we obtained the fluxes in the bands of Fiore's analysis for comparison. The
results are summarised in Figs.~\ref{fig:fluxcompa} and ~\ref{fig:fluxcompb}.

\begin{table}
\centering
\begin{tabular}{lllllll}
Source & N$_{H}^{a}$ &  $\Gamma$&$\chi^{2}_{\nu}$/d.o.f. &L$_{2-10}^{b}$\\  
ID&(10$^{21})$ & &&(10$^{44}$ ergs s$^{-1}$)\\   \hline
P1 & 0.0$^{^{+0.2}_{-0.0}}$ & 1.82$^{\pm 0.4}$ & 201 / 221&13\\
P2 & 0.0$^{^{+0.5}_{-0.0}}$ & 1.95$^{\pm 0.15}$& 88   /110&0.9\\
P3 & 4.7$^{^{+13}_{-4.7}}$ &  1.7$^{^{+2.9}_{-0.7}}$ &51   / 56&0.007\\
P4 & 1.3$^{^{+1.3}_{-1.0}}$ &1.67$^{^{+0.15}_{-0.13}} $ &56  /95&0.7\\
P5 &0.4 $^{^{+1.9}_{-0.4}}$ &1.88$^{^{+0.16}_{-0.17}}$&68   / 85&2.1\\
P6 & 0.2 $^{^{+0.4}_{-0.2}}$&1.95$^{^{+0.11}_{-0.12}}$&109  /146&0.46\\
 \label{tab:spect}
\end{tabular}
\caption[]{Summary of the source spectral parameters (90\% confidence ranges ). $^{a}$Additional absorption at source redshift. $^{b}$Assumes H$_{o}$=75 and q$_{o}$=0.5 }

 \end{table}

\section{Discussion}
\subsection{Individual Sources}
 For source P1 which is the brightest of the serendipitous sources, the
 XMM effective area would allow more detailed fitting of spectral details
 than heretofore on such a faint object.  However an acceptable fit is
 already evident with the baseline model.
 
 We were concerned initially about contamination of source P2 by the
 on-axis target. However we note that at $\Gamma$=2.4,  P2 has the steepest
 of  all  the  spectral slopes, yet $\Gamma$ for PKS0312-770 is 1.79, so
 that un-corrected  contamination would {\em harden} the spectrum
 compared with that actually obtained.  Since we reduce the
 accumulation radius from the 90\% encircled energy to 80\%,
 the energy-dependent part of the point-spread function caused by enhanced
 scattering at higher energies  could  result  in a steeper spectrum. However 
even assuming a 50\% 
 miscalibration of
 the  energy dependent inclusion fraction, the greatest change in slope would
 be $\sim $0.02. Thus we conclude P2 is truly a steep spectrum AGN source.
 
 For source P3, fixing the absorption to the galactic column provides a
 power law slope of 1.0 (comparable with the estimate of Fiore et
 al	\cite{Fiore} )  but  this  is an  unacceptable fit (Reduced $\chi ^{2} \sim$ 1.8). 
 The free absorbing column added at the source redshift is formally, the
 highest value in this sample and the resulting power  law  returned is typical for 
 an AGN, suggesting that the
 source is moderately absorbed, rather than demanding a more exotic
 emission mechanism. However neither the power law slope, nor the excess
 absorption are  well-constrained, rendering it difficult to speculate on
 the emission mechanisms in detail.
 
Source P4 was poorly characterised in the Fiore et al (\cite{Fiore})
optical spectroscopy. These authors speculate it is a Type 2 AGN with moderate
absorption. We find slight evidence to constrain the absorption
beyond the Galactic column, and find that the power law is harder
than most of the objects in the sample, tending towards a type 2 AGN.  

 Source P6 is close to the edge of the MOS CCDs so a smaller encircled 
energy was used, with a possible underestimate
 of the energy dependent correction factor.  However the spectral parameters
are consistent with those obtained from the PN camera, which has more  than  half  
 the combined effective area. Therefore encircled energy miscalibrations, if any,  
are expected to be small.  
 
\subsection{Flux Normalisation}
We find a slight evidence that the XMM-determined fluxes are in excess of the CHANDRA
estimated fluxes by about 10 (20)\% in the soft (hard) band. There is no systematic 
trend in this variation with spectral slope, off-axis
angle or brightness. As seen in Fig.~\ref{fig:fluxcompa}, the discrepancies
are small, and could be due to the normalisations resulting from the different models applied.
 As the XMM fluxes are estimated from a 90\%
encircled energy extraction region, a factor 20\% over-estimate is
thought unlikely to be due simply to the choice of extraction region. 
There is an on-going effort to reconcile the relative flux normalisations
between various observatories (XMM, CHANDRA, ROSAT, ASCA and Beppo-SAX) but to
date there is no evidence that XMM is significantly discrepant.

Fitting the MOS and PN cameras separately gives a flux determination on
the sources which is typically within 5 - 10\% consistent between the two
camera types, despite the different vignetting corrections and
 different instrument behaviours. This allows some measure of {\em systematic}
 errors in the XMM-Newton fluxes, and this is lower than
 the apparent discrepancy between XMM EPIC and CHANDRA. 
  
\section{Conclusions}
The increased XMM-Newton effective area allows a substantial improvement
in photon-gathering capability compared with CHANDRA. This has been used
in a follow-up of CHANDRA serendipitous sources. An unusual X-ray
luminous, but otherwise apparently normal galaxy (P3) seems to be most likely
explained as an obscured AGN. 

We report a small deviation in fluxes derived from
CHANDRA and XMM-Newton measurements. In the context of performing
systematic comparisons of Log N-Log S studies, for Sunyaev-Zeldovich
effect measurements, and source variability studies, this discrepancy in
flux normalisations should be addressed by dedicated comparisons of selected
spectral standard targets. 

The LogN-LogS curve derived from this XMM-Newton field is consistent
with population densities in the XMM-Newton Lockman Hole field, albeit limited
by the different exposure duration, to a higher flux range.

The diagnosis of source populations of this  exposure duration, typical
for that expected in many guest Observer programs, confirms the rich
serendipitous information to be obtained with XMM. Source fluxes can be measured down to $\sim$10$^{-15}$
ergs cm$^{-2}$ s$^{-1}$, and hardness ratios determined to identify particularly
extreme objects.
\begin{acknowledgements}

We thank the members of the XMM Science Operations Centre, particularly
those whose efforts in mission planning and operations activities were
so crucial to the execution of the Calibration / PV phases of the
mission.

\end{acknowledgements}

\end{document}